\documentclass[a4paper]{jpconf}
\usepackage{graphicx}

\newcommand{\be}{\beta}
\newcommand{\ga}{\gamma}

\newcommand{\De}{\Delta}

\newcommand{\La}{\Lambda}
\newcommand{\om}{\omega}
\newcommand{\sig}{\sigma}

\def\cL{{\cal L}} \def\cM{{\cal M}}

\newcommand{\unity}{1\hspace{-0.15cm}1}
\newcommand{\nn}{\nonumber}
\newcommand{\mean}[1]{\langle#1\rangle}

\def\dd{\displaystyle}

\newcommand{\beq}{\begin{equation}}
\newcommand{\eeq}{\end{equation}}
\newcommand{\bac}{\beq\begin{array}}
\newcommand{\eac}{\end{array}\eeq}
\newcommand{\ba}{\begin{array}}
\newcommand{\ea}{\end{array}}
\newcommand{\bea}{\begin{eqnarray}}
\newcommand{\eea}{\end{eqnarray}}

\renewcommand{\Im}{\mathrm{Im}}
\renewcommand{\Re}{\mathrm{Re}}
\newcommand{\phit}{\varphi_T}
\newcommand{\phis}{\varphi_S}
\newcommand{\onep}{1^\prime}
\newcommand{\onepp}{1^{\prime\prime}}

\begin{document}
\title{LFV and Dipole Moments in Models with A4 Flavour Symmetry}

\author{Luca Merlo}

\address{Dipartimento di Fisica `G.~Galilei', Universit\`a di Padova \&
INFN, Sezione di Padova, Via Marzolo~8, I-35131 Padova, Italy}

\ead{merlo@pd.infn.it}

\begin{abstract}
It is presented an analysis on lepton flavour violating transitions, leptonic magnetic dipole moments and electric dipole moments in a class of models characterized by the flavour symmetry $A_4\times Z_3\times U(1)_{FN}$, whose choice is motivated by the approximate Tri-Bimaximal mixing observed in neutrino oscillations. A low-energy effective Lagrangian is constructed, where these effects are dominated by dimension six operators, suppressed by the scale M of new physics. All the flavour breaking effects are universally described by the vacuum expectation values $\mean{\Phi}$ of a set of spurions. Two separate cases, a supersymmetric and a general one, are described. An upper limit on $\theta_{13}$ of a few percent is concluded.
\end{abstract}


\section{Introduction}
It is a common belief that the solar and the atmospheric neutrino anomalies can be explained by the neutrino oscillations. In table \ref{table:OscillationData}, we can read the results of two independent global fits to neutrino oscillation data from \cite{Fogli_Data} and \cite{Schwetz_Data}.

\begin{table}[ht]
\caption{\label{table:OscillationData} Neutrino oscillation parameters from independent global fits \cite{Fogli_Data, Schwetz_Data}.}
\begin{center}
\begin{tabular}{lcccc}
\br
& \multicolumn{2}{c}{Ref.~\cite{Fogli_Data}} & \multicolumn{2}{c}{Ref.~\cite{Schwetz_Data}}\\
parameter & best fit(\@$1\sig$) & 3$\sig$-interval & best fit(\@$1\sig)$ & 3$\sig$-interval\\
\mr
$\De m^2_{21}\:[10^{-5}\mathrm{eV}^2]$
        & $7.67^{+0.16}_{-0.19}$ & $7.14-8.19$
        & $7.65^{+0.23}_{-0.20}$ & $7.05-8.34$\\[2mm]
$|\De m^2_{31}|\: [10^{-3}\mathrm{eV}^2]$
        & $2.39^{+0.11}_{-0.8}$ & $2.06-2.81$
        & $2.40^{+0.12}_{-0.11}$ & $2.07-2.75$\\[2mm]
$\sin^2\theta_{12}$
        & $0.312^{+0.019}_{-0.018}$ & 0.26-0.37
        & $0.304^{+0.022}_{-0.016}$ & 0.25-0.37\\[2mm]
$\sin^2\theta_{23}$
        & $0.466^{+0.073}_{-0.058}$ & 0.331-0.644
        & $0.50^{+0.07}_{-0.06}$ & 0.36-0.67\\[2mm]
$\sin^2\theta_{13}$
        & $0.016^{+0.010}_{-0.010}$ & $\leq$ 0.046
        & $0.010^{+0.016}_{-0.011}$ & $\leq$ 0.056\\
\br
\end{tabular}
\end{center}
\end{table}

The pattern of the mixing angles is quite particular, indeed two of them are large while the third is extremely small: the atmospheric angle $\theta_{23}$ is compatible with a maximal value, but the accuracy admits relatively large deviations; the solar angle $\theta_{12}$ is large, but about $5\sigma$ errors far from the maximal value; the reactor angle $\theta_{13}$ only has an upper limit. We underline that there are contrasting indications for a vanishing value of the reactor angle: in \cite{Fogli_Data} there is a suggestion for a positive value which, at $1.6\sig$, is $\sin^2\theta_{13}\simeq0.016\pm0.010$, while in \cite{Maltoni_Indication} the authors find a best fit value consistent with zero within less than $1\sig$. Therefore we need a confirmation by the future experiments like DOUBLE CHOOZ,  Daya Bay and MINOS in the $\nu_e$ appearance channel \cite{Niki}.

From the theoretical point of view, we are experiencing the lack of a unique and compelling picture which provides a natural explanation of the neutrino physics. However a series of models based on some discrete non-Abelian groups seems to be extremely attractive due to their predictions and naturalness: indeed it is possible to achieve as the neutrino mixing matrix the Tri-Bimaximal (TB) pattern \cite{TB1,TB2},
\beq
U_{TB}=\left(
         \begin{array}{ccc}
           \sqrt{2/3} & 1/\sqrt3 & 0 \\
           -1/\sqrt6 & 1/\sqrt3 & -1/\sqrt2 \\
           -1/\sqrt6 & 1/\sqrt3 & +1/\sqrt2 \\
         \end{array}
       \right)\;,
\eeq
which represents a very good approximation of the experimental data in table \ref{table:OscillationData}:
\beq
\sin^2\theta_{13}^{TB}=0\qquad\sin^2\theta_{23}^{TB}=1/2\qquad\sin^2\theta_{12}^{TB}=1/3\;.
\eeq

We get fundamental indications for the symmetry that best describes the neutrino mixings just from the TB pattern: it is a very well known result \cite{AF_ED} that a maximal value for the atmospheric angle can be recovered only with a non exact symmetry; explaining the indication for a non vanishing, but still very small, value for $\theta_{13}$, it is necessary to provide the TB pattern at the leading order (LO), invoking corrections from the higher order terms; the solar angle is predicted to be very close, less than $2^\circ$, to the measured best value and therefore the corrections has to be relatively small. As a result, a realistic lepton flavour symmetry has to be broken at a certain level, predicting at the LO the TB pattern and providing corrections at the next-to-the-leading order (NLO) at most of the order of $\theta_c^2\approx2^\circ$, where $\theta_c$ stands for the Cabibbo angle, which is a convenient hierarchical parameter for both the sectors.

There is a series of models based on the symmetry group $A_4$ \cite{TBA4_General}, which are extremely attractive from this point of view, fulfilling all the previous requirements (recent alternatives have been proposed using different flavour groups as \cite{King,BMMS4,Ivo}). These models manage in deriving the TB mixing by assuming that the $A_4$ symmetry is realized at a very high energy scale $\La$ and that leptons transform in a non trivial way under this symmetry. Afterward the group is spontaneously broken by a set of scalar multiplets $\Phi$, the flavons, whose vacuum expectation values (VEVs) receive a specific alignment. The realization of the required vacuum alignment is a non trivial task and we can find a natural explanation of it in \cite{AF_ED,AF_Modular,AFL_Orbifold}, which are our main references and we referee to this model as the Altarelli-Feruglio (AF) model (an alternative but still interesting approach has been provided in \cite{Lin_Predictive}). The TB mixing is corrected by the higher order terms by quantities of the order of $\mean{\Phi}/\La<1$ and as a result the reactor angle is no longer vanishing and becomes proportional to $\mean{\Phi}/\La$.

The predictions for the lepton mixings and for the spectrum of the AF model are not sufficient to distinguish it from all the other models, which present the TB pattern. For this reason, we looked \cite{FHLM_LFV} for other observables, not connected to the neutrino oscillations, which can describe with more details the AF model. This is possible if we introduce an intermediate energy scale $M$ at about $1\div10$ TeV: this corresponds to the presence of some kind of new physics (NP) at this scale. Other indications, which enforce this choice, come from for example the discrepancy in the anomalous magnetic moment of the muon, the presence of Dark Matter, the convergence to a unique value of the gauge coupling constants and the solution to the hierarchy problem, which all would benefit by the presence of NP at $1\div10$ TeV.\\

In the following, we present the AF model in section \ref{section:AF_Model} and then, in section \ref{section:EFT}, we analyze the predictions of the model for a set of relevant low-energy observables, as lepton flavour violating (LFV) transitions, leptonic magnetic dipole moments (MDM) and electric dipole moments (EDM). In both the sections we separate the discussion among two distinct cases, a supersymmetric and a general one. The results are briefly summarized in section \ref{section:Conclusions}.


\section{The Altarelli-Feruglio Model}
\label{section:AF_Model}

$A_4$ is the group of even permutations of four objects, isomorphic to the group of discrete rotations in the three-dimensional space that leave invariant a regular tetrahedron. It is generated by two elements
$S$ and $T$ obeying the relations $S^2=(ST)^3=T^3=\unity$.
The group $A_4$ has two obvious subgroups: $G_S$ isomorphic to $Z_2$, generated by $S$, and $G_T$ isomorphic to $Z_3$, generated by $T$. They are the relevant low-energy symmetries of the neutrino and the charged-lepton sectors at LO, respectively. The TB mixing is then a direct consequence of this special symmetry breaking pattern, which is achieved via the vacuum misalignment of triplet scalar fields, called flavons.
\begin{table}[ht]
\caption{\label{table:transformations} The transformation rules of the fields, following \cite{AF_Modular,FHLM_LFV}.}
\begin{center}
\begin{tabular}{cccccccccc}
\br
 & $\ell$ & $e^c$ & $\mu^c$ & $\tau^c$ & $H$ & $\phit$ & $\phis$ & $\xi$ & $\theta$\\
\mr
$A_4$ & 3 & 1 & $\onepp$ & $\onep$ & 1 & 3 & 3 & 1 &  1 \\
$Z_3$ & $\om$ & $\om^2$ & $\om^2$ & $\om^2$  & 1 & 1& $\om$ & $\om$  & 1 \\
$U(1)$ & 0 & 2 & 1 & 0 & 0 & 0 & 0 & 0 & -1  \\
\br
\end{tabular}
\end{center}
\end{table}

Concerning the complete flavour group, following \cite{AF_ED,AF_Modular}, it is chosen to be
\beq
A_4\times Z_3\times U(1)_{FN}\;,
\eeq
where the three factors play different roles: the spontaneous breaking of $A_4$ is directly responsible for the TB mixing; the $Z_3$ factor has a similar behaviour as the continuous total lepton number; the $U(1)_{FN}$ is responsible for the hierarchy among the charged fermion masses.
The transformation properties of the lepton fields $\ell$, $e^c$, $\mu^c$, $\tau^c$, of the electroweak scalar doublet $H$ and of the flavon fields are summarized in table \ref{table:transformations}.
The Yukawa interactions in the lepton sector at the LO read:
\bea
{\cal L}_\ell^{LO}&=&\dd\frac{y_e}{\Lambda^3} \theta^2e^c H^\dagger \left(\varphi_T \ell\right)
+\dd\frac{y_\mu}{\Lambda^2} \theta\mu^c H^\dagger \left(\varphi_T \ell\right)'
+\dd\frac{y_\tau}{\Lambda} \tau^c H^\dagger \left(\varphi_T \ell\right)''\label{Ll}+\\
&&+\dd\frac{y_e'}{\Lambda^3} \theta^2e^c H^\dagger \left(\varphi_T^\dagger \ell\right)
+\dd\frac{y_\mu'}{\Lambda^2} \theta\mu^c H^\dagger \left(\varphi_T^\dagger \ell\right)'
+\dd\frac{y_\tau'}{\Lambda} \tau^c H^\dagger \left(\varphi_T^\dagger \ell\right)''+h.c.\nn\\
\nn\\
{\cal L}_\nu^{LO}&=&
\dd\frac{x_a}{\Lambda^2} \xi ({\tilde H}^\dagger \ell {\tilde H}^\dagger \ell) +\dd\frac{x_b}{\Lambda^2} (\varphi_S {\tilde H}^\dagger \ell {\tilde H}^\dagger \ell)+h.c.
\label{lnu}
\eea
The flavour symmetry breaking sector of the model includes the scalar fields $\phit$, $\phis$, $\xi$ and $\theta$:
\beq
\langle\dd\frac{\phit}{\La}\rangle=(u,0,0)\quad\langle\dd\frac{\phis}{\La}\rangle=c_b(u,u,u)\quad\langle\dd\frac{\xi}{\La}\rangle=c_a u\quad\langle\dd\frac{\theta}{\La}\rangle=t\;,
\label{VEV:alignment}
\eeq
where $c_a$ and $c_b$ are constants of order 1, while $u$ and $t$ are the small symmetry breaking parameters of the theory, which parameterize the ratio of the VEVs of the flavons over the cut-off $\La$. It has been introduced a different parameter $t$ for $\mean{\theta}$, because it has a different origin with respect to the other VEVs. The VEV misalignment in eq.(\ref{VEV:alignment}) comes from the minimization of the scalar potential and a complete proof can be found in \cite{AF_ED,FHLM_LFV,AFH_SU5xA4}. With this setting, the model predicts, at the LO, a diagonal charged lepton mass matrix
\beq
m_\ell=\left(
         \begin{array}{ccc}
           \hat{y}_e t^2 & 0 & 0 \\
           0 & \hat{y}_\mu t & 0 \\
           0 & 0 & \hat{y}_\tau \\
         \end{array}
       \right)uv\;,
\eeq
where $v\approx 174$ GeV is the electroweak breaking scale, defined as $1/\sqrt{2}$ times the VEV of the real, electrically neutral, component of $H$ and
\beq
u=|u| \, \mathrm{e}^{i \psi}\qquad\hat{y}_{f} = y_{f} + y_{f} ^{\prime} \mathrm{e}^{-2 i \psi}\qquad(f=e,\mu,\tau)\;.
\eeq
Looking at $m_\ell$, it is possible to get some bounds on the parameters $u$ and $t$. A lower bound on $|u|$ of about $0.001$ comes from the requirement that the Yukawa coupling of the $\tau$ remains in the perturbative regime, while from the hierarchy between the charged lepton masses,
\beq
\dd\frac{|m_e|}{|m_\mu|}=\dd\frac{|m_\mu|}{|m_\tau|}\sim |t|\;,
\eeq
$|t|$ has to be of about $0.05$, value which can be parameterized by $\theta_c^2$. Furthermore, the model predicts a neutrino mass matrix which is exactly diagonalized by the TB mixing matrix
\beq
m_\nu=\left(
        \begin{array}{ccc}
            a+2 b/3& -b/3& -b/3\\
            -b/3& 2b/3& a-b/3\\
            -b/3& a-b/3& 2 b/3
        \end{array}
        \right)\dd\frac{v^2}{\La}\;,
\eeq
where $a\equiv 2 x_ac_au$ and $b\equiv 2 x_bc_bu$. The leading order predictions are $\tan^2\theta_{23}=1$, $\tan^2\theta_{12}=0.5$ and $\theta_{13}=0$. The neutrino masses e $m_1=a+b$, $m_2=a$ and $m_3=-a+b$, in units of $v_u^2/\Lambda$. In \cite{AF_ED} it has been proven that only a Normal Hierarchy (NH) or a Quasi Degenerate (QD) spectrum can be explained and that it is needed a moderate fine-tuning in order to provide the correct value for $r\equiv \Delta m^2_{sol}/\Delta m^2_{atm}$. In the case of the NH, there are a series of interesting bounds for the lightest neutrino mass, $m_{\nu1}>13.8$ meV, the sum of neutrino masses, $\sum m_{\nu i}>77.2$ meV, and for $|m_{ee}|$, the parameter characterizing the violation of total lepton number in $0\nu2\be$ decay, which is $|m_{ee}|>3.7$ meV, at the upper edge of the range allowed for the NH and not too small to be detected in the near future.
Independently from the type of the spectrum, it is present this relation
\beq
|m_3|^2=|m_{ee}|^2+\dd\frac{10}{9}\Delta m^2_{atm}\left(1-\dd\frac{r}{2}\right)\;,
\eeq
which is a prediction of the model.\\

When the NLO terms are considered, there are two kind of corrections. The first type comes from the VEV of the flavons, which become at the NLO
\beq
\langle\dd\frac{\phit}{\La}\rangle=(u+c_1u^2,c_2u^2,c_3u^3)\qquad\qquad \langle\dd\frac{\phis}{\La}\rangle=c_b(u,u,u)+O(u^2)\;,
\label{VEV:alignmentNLO}
\eeq
where it is not interesting to explicit the correction of $\mean{\phis}$, while that one of $\mean{\xi}$ can be reabsorbed in $c_a$. As a result the operators of eq.(\ref{Ll}) with these insertions for the VEVs give rise to deviations of relative order $u$ with respect to the LO results. The second source of NLO corrections consists in the higher order operators in the Lagrangian $\cL_\ell^{LO}$, which we can write as
\beq
{\cal L}_\ell^{NLO}=\sum_{i=1}^6\dd\frac{y_e^i}{\Lambda^4} \theta^2e^c H^\dagger \left(X_i \ell\right)
+\sum_{i=1}^6\dd\frac{y_\mu^i}{\Lambda^3} \theta\mu^c H^\dagger \left(X_i \ell\right)'
+\sum_{i=1}^6\dd\frac{y_\tau^i}{\Lambda^2} \tau^c H^\dagger \left(X_i \ell\right)''+h.c.
\label{LlNLO}
\eeq
where
\beq
X=\left\{\varphi_T^2,(\varphi_T^\dagger)^2,\varphi_T^\dagger\varphi_T, \varphi_S^\dagger\varphi_S,\xi^\dagger\varphi_S,\varphi_S^\dagger\xi\right\}\;.
\label{Xlist}
\eeq
Only some of the terms in the list of eq.(\ref{Xlist}) give non trivial corrections: the first three can be reabsorbed in the leading order terms; the fourth one is vanishing; the last two give genuine contributions. Also these last terms give rise to deviations of relative order $u$ and the corrected charged lepton mass matrix appears, in terms of order of magnitude, as
\begin{equation}
m_\ell= \left(
               \begin{array}{ccc}
                 O(t^2 u) & O(t^2 u^2) & O(t^2 u^2) \\
                 O(t u^2) & O(t u) & O(t u^2) \\
                 O(u^2) & O(u^2) & O(u) \\
               \end{array}
             \right)uv\;.
\label{ml_subleading}
\end{equation}
A similar discussion yields for the neutrino sector, but the subleading Lagrangian does not have a particular flavour structure and then it is sufficient to report that the corrections are of relative order $u$. The main result is that the predicted neutrino mixing angles are modified by factors of order $u$ and therefore
\beq
\theta_{13}=O(u)\qquad\theta_{23}=\dd\frac{\pi}{4}+O(u)\;,
\eeq
which is extremely useful to account for a positive value of the reactor angle. A second important consequence is that the corrections cannot be too large, in order to preserve the agreement between the TB value for $\theta_{12}$ and the experimental measure. An upper bound on $|u|$ of about $\theta_c^2$ has to be imposed.

\subsection{The Supersymmetric Case}
It has been presented in \cite{AF_Modular} a supersymmetric version of the model, which possesses different and interesting predictions for some LFV transitions, that we will discuss in the next section.

In the supersymmetric case, the transformation properties listed in table \ref{table:transformations} are still valid, two Higgs doublets $h_{u,d}$, invariant under $A_4$, are introduced together to an additional flavon field $\xi'$ which transforms exactly as $\xi$, but with vanishing VEV. The Lagrangian $\cL_\ell^{LO}$ of eq.(\ref{Ll}) is identified to the superpotential $w_\ell^{LO}$
\bea
w_\ell^{LO}&=&\dd\frac{y_e}{\La^3}\theta^2e^ch_d\left(\phit\ell\right)+
\dd\frac{y_\mu}{\La^2}\theta\mu^ch_d\left(\phit\ell\right)'+
\dd\frac{y_\tau}{\La}\tau^ch_d\left(\phit\ell\right)''+h.c.
\label{wl}\\
w_\nu^{LO}&=&\dd\frac{x_a}{\La^2}\xi\left(h_u\ell h_u\ell\right)+
\dd\frac{x_b}{\La^2}\left(\phis h_u\ell h_u\ell\right)+h.c.\;.
\label{wnu}
\eea
It is fundamental to underline that only the terms with the flavon $\phit$ are present in the superpotential, while those with $\phit^\dag$ are forbidden in the supersymmetric context.
The VEVs of the flavons are the same to the previous case only at the LO, but considering also the NLO terms the corrections to the second and third entries of $\mean{\phit}$ become equal and finally we have
\beq
\langle\dd\frac{\phit}{\La}\rangle=(u+c_1u^2,c_2u^2,c_2u^3)\qquad \langle\dd\frac{\phis}{\La}\rangle=c_b(u,u,u)+O(u^2)\qquad \langle\dd\frac{\xi}{\La}\rangle=c_a u\qquad \langle\dd\frac{\theta}{\La}\rangle=t\;.
\label{VEV:alignmentNLOSUSY}
\eeq
When considering the higher order terms in the superpotential, only few operators are admitted in comparison with eq.(\ref{LlNLO}), due to the supersymmetric context (for more details see \cite{FHLM_LFV}):
\beq
w_\ell^{NLO}=\dd\frac{y'_e}{\Lambda^4} \theta^2e^c h_d \left(\phit\phit\ell\right)
+\sum_{i=2}^7\dd\frac{y'_\mu}{\Lambda^3} \theta\mu^c h_d \left(\phit\phit\ell\right)'
+\sum_{i=2}^7\dd\frac{y'_\tau}{\Lambda^2} \tau^c h_d \left(\phit\phit\ell\right)''+h.c.\;.
\label{wlNLO}
\eeq
All these terms give contributions only to the diagonal entries and can be reabsorbed in the LO parameters $y_i$; the only genuine corrections to the LO result come from the deviations of the vacuum alignment of the flavons. The final charged lepton mass matrix presents the same order of magnitude in the single entries as in eq.(\ref{ml_subleading}), but with a more simple structure.

All the predictions illustrated in the general non supersymmetric case are still valid. Furthermore, the case with the see-saw mechanism has been studied in \cite{AF_Modular}, while some extensions in order to describe quarks have been proposed in \cite{AFH_SU5xA4,FHLM_Tprime}.


\section{Effective Operator Analysis}
\label{section:EFT}

In this section we discuss the analysis performed in \cite{FHLM_LFV} on some LFV transitions, like $\mu\to e\gamma$, $\tau\to\mu\gamma$ and $\tau\to e\gamma$, and on the lepton EDMs and MDMs, in order to find new characterizing feature of the AF model, useful to discriminate it among all the models with similar predictions for observables linked to the neutrino oscillations. We start dealing with the non supersymmetric version of the AF model. This finds a well defined reason in the type of NP, which we assume to exist at the energy scale $M$, at about $1\div10$ TeV: we consider the Standard Model context and the presence of some kind of NP, without specifying it at this level, only requiring that the new degrees of freedom do not provide new sources of baryon and/or lepton number violation. Following this choice, we perform an effective field theory approach, which is ideal in this case because it does not require to specify the spectrum above a certain energy level. We first integrate out the d.o.f. related to $\La$ and then those related to $M$: the dominant physical effects of the NP at low energies
can be described by dim-6 operators, suppressed by $M^2$ and not by $\Lambda^2$, opening the possibility that the related effects might be seen in the near future.

The leading terms of the relevant effective Lagrangian are
\begin{equation}
\cL_{eff}=\cL_{SM}+\delta \cL(m_\nu)+i\dd\frac{e}{M^2} {e^c}^T H^\dagger \sigma F \cM\ell+\ldots
\label{Leff}
\end{equation}
where $e$ is the electric charge, $e^c$ the set of SU(2) lepton singlets,
$F_{\mu\nu}$ is the electromagnetic field strength and $\cM\equiv\cM\left(\langle\phi\rangle\right)$ is a complex $3\times 3$ matrix $\cM$, with indices in the flavour space. In the basis of canonical kinetic terms and diagonal charged leptons (we denote by a hat the relevant matrices in this basis), the real and imaginary parts of the matrix elements $\hat{\cal M}_{ii}$ are proportional to the MDMs $a_i$ and to the EDMs $d_i$ of the charged leptons, respectively:
\beq
a_i=2 m_i \dd\frac{v}{\sqrt{2} M^2}Re \hat{\cal M}_{ii}\qquad d_i=e \dd\frac{v}{\sqrt{2} M^2}Im \hat{\cal M}_{ii}\qquad(i=e,\mu,\tau)\;.
\eeq
The off-diagonal elements $\hat{\cal M}_{ij}$ describe the amplitudes for the LFV transitions $\mu\to e \gamma$, $\tau\to\mu\gamma$ and $\tau\to e \gamma$:
\beq
\dd\frac{BR(\ell_i\to \ell_j\gamma)}{BR(\ell_i\to \ell_j\nu_i{\bar \nu_j})}=\dd\frac{12\sqrt{2}\pi^3 \alpha}{G_F^3 m_i^2 M^4}\left(\vert\hat{\cal M}_{ij}\vert^2+\vert\hat{\cal M}_{ji}\vert^2\right)
\eeq
where $\alpha$ is the fine structure constant, $G_F$ is the Fermi constant and $m_i$ is the mass of the lepton $\ell_i$. We can say that all the new observables we are interested in are connected to the dipole matrix $\hat{\cM}$ and therefore it is a fundamental point to evaluate it.

In order to reach this result, we should analyze the dimension six operators which give rise to the third factor in the right-hand part of eq.(\ref{Leff}): we call all these terms with ${\cal L}_{dip}$
\beq
{\cal L}_{dip}={\cal L}_{dip}^{LO}+{\cal L}_{dip}^{NLO}+\ldots
\eeq
with
\bea
\label{Ld}
{\cal L}_{dip}^{LO}&=&i\dd\frac{e}{M^2}\left[\dd\frac{\beta_e}{\Lambda^3} \theta^2e^c H^\dagger \sigma\cdot F \left(\varphi_T \ell\right)
+\dd\frac{\beta_\mu}{\Lambda^2} \theta\mu^c H^\dagger \sigma\cdot  F \left(\varphi_T \ell\right)'
+\dd\frac{\beta_\tau}{\Lambda} \tau^c H^\dagger \sigma\cdot F \left(\varphi_T \ell\right)''+\right.\\
&+&\left.\dd\frac{\beta_e'}{\Lambda^3} \theta^2e^c H^\dagger \sigma\cdot F \left(\varphi_T^\dagger \ell\right)
+\dd\frac{\beta_\mu'}{\Lambda^2} \theta\mu^c H^\dagger \sigma\cdot  F \left(\varphi_T^\dagger \ell\right)'
+\dd\frac{\beta_\tau'}{\Lambda} \tau^c H^\dagger \sigma\cdot F \left(\varphi_T^\dagger \ell\right)''\right]+h.c.\nonumber
\eea
and
\beq
{\cal L}_{dip}^{NLO}=i\dd\frac{e}{M^2}\sum_{i=1}^6\left[\dd\frac{\beta_e^i}{\Lambda^4} \theta^2e^c H^\dagger \sigma\cdot F \left(X_i l\right)
+\dd\frac{\beta_\mu^i}{\Lambda^3} \theta\mu^c H^\dagger \sigma\cdot  F \left(X_i \ell\right)'
+\dd\frac{\beta_\tau^i}{\Lambda^2} \tau^c H^\dagger \sigma\cdot F \left(X_i \ell\right)''\right]+h.c.
\label{LdNLO}
\eeq
Comparing these expressions with eqs.(\ref{Ll}) and (\ref{LlNLO}), we note that they are controlled by the same symmetry breaking parameters and we get that $\cM$ has the same flavour structure of the charged lepton mass matrix of eq.(\ref{ml_subleading}): we only report here the order of magnitude of the expression found for $\cM$ (for the complete discussion see \cite{FHLM_LFV})
\begin{equation}
\mathcal{M} = \left(
               \begin{array}{ccc}
                 O(t^2 u) & O(t^2 u^2) & O(t^2 u^2) \\
                 O(t u^2) & O(t u) & O(t u^2) \\
                 O(u^2) & O(u^2) & O(u) \\
               \end{array}
             \right)\;.
\label{M_subleading}
\end{equation}

It is possible now to compute the physical observable quantities we are interested in, MDMs, EDMs and flavour violating transitions for leptons. Working at the lowest order in the symmetry breaking parameter $u$, the kinetic terms are canonical and $m_\ell$ is diagonal. In this approximation also the dipole matrix $\mathcal{M}$
is diagonal and it only contributes to lepton MDMs and EDMs, which we will discuss below. In order to find also the size of the flavour violating processes we need to include the sub-leading effects originating from insertions of the flavon fields $\varphi_{S,T}$ and $\xi$, shifts of the VEV of $\varphi_T$ and (additional) insertions of $\theta$ in the Lagrangian. These generate non-canonically normalized kinetic terms and render $m_\ell$ and $\mathcal{M}$ non diagonal, like in eqs.(\ref{ml_subleading}) and (\ref{M_subleading}). Therefore, we first have to move to the basis where the kinetic terms are canonically normalized and the charged lepton mass matrix is diagonal: the resulting dipole matrix preserves the same flavour structure as in eq.(\ref{M_subleading}), but the coefficients of the single entries change.

The MDMs and EDMs are given at lowest order as:
\beq
\begin{array}{rll}
a_e &= 2 m_e\dd\frac{v}{\sqrt{2} M^2} \Re{[O(t^2\,u)]}\\[3mm]
a_\mu &= 2 m_\mu\dd\frac{v}{\sqrt{2} M^2} \Re{[O(t\,u)]}\\[3mm]
a_\tau &= 2 m_\tau\dd\frac{v}{\sqrt{2} M^2} \Re{[O(u)]}
\end{array}\qquad
\begin{array}{rll}
d_e&=e\dd\frac{v}{\sqrt{2} M^2} \Im{[O(t^2\,u)]}\\[3mm]
d_\mu&=e\dd\frac{v}{\sqrt{2} M^2} \Im{[O(t\,u)]}\\[3mm]
d_\tau&=e\dd\frac{v}{\sqrt{2} M^2} \Im{[O(u)]}\;.
\end{array}
\eeq
Notice that we can write these expressions in function of the charged leptons masses:
\beq
a_i\sim O\left(2\dd\frac{m_i^2}{M^2}\right)\qquad d_i\sim O\left(e\dd\frac{m_i}{M^2}\right)\;.
\label{DM}
\eeq

\begin{table}[ht]
\caption{\label{table:MfromMDMEDM}Experimental limits on lepton MDMs and EDMs and corresponding bounds on the scale $M$, derived from eqs.(\ref{DM}).
The data on the $\tau$ lepton have not been reported since they are much less constraining. For the anomalous magnetic moment of the muon,
$\delta a_\mu$ stands for the deviation of the experimental central value from the SM expectation.}
                \begin{center}
                \begin{math}
                \begin{array}{ll}
                    \br
                    d_e<1.6\times 10^{-27}\textrm{ e\,cm}&\qquad M>80\textrm{ TeV}\\
                    d_\mu<2.8\times 10^{-19}\textrm{ e\,cm}&\qquad M>80\textrm{ GeV}\\
                    \mr
                    \delta a_e<3.8\times 10^{-12}&\qquad M>350\textrm{ TeV}\\
                    \delta a_\mu\approx 30\times 10^{-10}&\qquad M\approx 2.7\textrm{ TeV}\\
                    \br
                \end{array}
               \end{math}
               \end{center}
\end{table}

We can derive a bound on the scale $M$, by considering the existing limits on MDMs and EDMs and by using the expressions in eqs.(\ref{DM}) as exact equalities: the results are listed in table \ref{table:MfromMDMEDM}. The strongest constraint, $M>80$ TeV, comes from the EDM of the electron, $d_e$: in order to lower this value in the range we have previously indicated, we need to invoke a cancellation in $\Im[\hat{\cal{M}}_{ee}]$, which could be accidental or due to some kind of CP conservation. Furthermore, a very interesting indication for LHC comes from $\delta a_\mu$, $M\approx2.7$ TeV.

Concerning the flavour violating dipole transitions, we see that the rate for $\ell_i\to \ell_j\gamma$ is dominated by
the contribution of ${\hat{\cal M}}_{ij}$:
\beq
\dd\frac{BR(\ell_i\to \ell_j\gamma)}{BR(\ell_i\to \ell_j\nu_i{\bar \nu_j})}=\dd\frac{48\pi^3 \alpha}{G_F^2 M^4}|w_{ij}\,u|^2\;.
\label{LFV}
\eeq
where $w_{ij}$ are coefficients of order 1.
The branching ratios for the three transitions $\mu\to e\gamma $, $\tau\to\mu\gamma$ and $\tau\to e\gamma$ are all of the same order:
\beq
BR(\mu\to e \gamma)\approx BR(\tau\to\mu\gamma)\approx BR(\tau\to e \gamma)\,.
\eeq
This is a distinctive feature of this models, since in most of the
other existing models there is a substantial difference between the branching ratios (consider for example the MFV models \cite{MFV,MLFV}).
Imposing the present experimental bound $BR(\mu\to e \gamma)<1.2\times 10^{-11}$, our result implies that $\tau\to\mu\gamma$ and $\tau\to e \gamma$ have rates much below the present and expected future sensitivity.  Furthermore, from the current (future) experimental limit on $BR(\mu\to e \gamma)$ and considering that the absolute value of $u$ lies in the limited range $0.001<u<0.05$, we find
\bea
\vert u\vert =& 0.001&\qquad M>10~(30)~~{\rm TeV}\\
\vert u\vert =& 0.05&\qquad M>70~(200)~~{\rm TeV}\;.
\eea
This pushes the scale $M$ considerably above the range we were initially interested in. In particular $M$ is shifted above the region of
interest for $(g-2)_\mu$ and probably for LHC.

\subsection{The Supersymmetric Case}
\label{subsection:SUSY_Case}
As we have seen, the off-diagonal elements of the dipole matrix ${\cal M}$ can be traced back to two independent sources.
They can originate either from ${\cal L}_{dip}^{LO}$, when the sub-leading corrections to the VEV of the $\varphi_T$
multiplet (terms proportional to $c_{1,2,3}$ in eq.(\ref{VEV:alignmentNLO})) are accounted for, or from ${\cal L}_{dip}^{NLO}$,
where the relevant double flavon insertions (see eq.(\ref{Xlist})) are considered. In this second case, the only combinations of flavon insertions that can provide non-vanishing contributions are $\xi^\dagger \varphi_S$ and its hermitian conjugate. In a generic case we expect that both these contributions are equally important and contribute at the same order to a given off-diagonal dipole transition. There is however a special case where the double flavon insertions $\xi^\dagger \varphi_S$ and its hermitian conjugate are suppressed compared to the sub-leading corrections to $\varphi_T$ and an overall depletion in the elements of the matrix $\hat{\cal M}$ below the diagonal takes place. This happens when the underlying theory is supersymmetric and supersymmetry is softly broken.

In this new context, the dipole matrix $\cM$ is given by the superpotential
\beq
w_{dip}=w_{dip}^{LO}+w_{dip}^{NLO}
\eeq
where
\beq
\label{wd}
w_{dip}^{LO}=i\dd\frac{e}{M^2}\left[\dd\frac{\beta_e}{\Lambda^3} \theta^2e^c h_d \sigma\cdot F \left(\varphi_T \ell\right)
+\dd\frac{\beta_\mu}{\Lambda^2} \theta\mu^c h_d \sigma\cdot  F \left(\varphi_T \ell\right)'
+\dd\frac{\beta_\tau}{\Lambda} \tau^c h_d \sigma\cdot F \left(\varphi_T \ell\right)''\right]+h.c.
\eeq
and
\beq
w_{dip}^{NLO}=i\dd\frac{e}{M^2}\left[\dd\frac{\beta'_e}{\Lambda^4} \theta^2e^c h_d \sigma\cdot F \left(\phit\phit l\right)
+\dd\frac{\beta'_\mu}{\Lambda^3} \theta\mu^c h_d \sigma\cdot  F \left(\phit\phit \ell\right)'
+\dd\frac{\beta'_\tau}{\Lambda^2} \tau^c h_d \sigma\cdot F \left(\phit\phit \ell\right)''\right]+h.c.
\label{wdNLO}
\eeq
All the terms in eq.(\ref{wdNLO}) give contributions only to the diagonal entries and can be reabsorbed in the parameters $\beta_i$; the only genuine corrections to the LO dipole matrix come from the deviations of the vacuum alignment of the flavon $\phit$. The resulting dipole matrix at NLO has the same structure as in eq.(\ref{M_subleading}), namely the entries present the same order of magnitude. Moving to the basis of canonical kinetic terms and diagonal charged lepton mass matrix, we get the main difference between the supersymmetric and the general context: here we get an overall depletion of order $u$ in the elements of the matrix $\hat{\cal M}$ below the diagonal and as a result
\begin{equation}
\mathcal{M} = \left(
               \begin{array}{ccc}
                 O(t^2 u) & O(t^2 u^2) & O(t^2 u^2) \\
                 O(t u^3) & O(t u) & O(t u^2) \\
                 O(u^3) & O(u^3) & O(u) \\
               \end{array}
             \right)\;.
\end{equation}
The EDMs and MDMs are similar to those of the general non-supersymmetric case: we find the same degree of suppression in the $t$ and $u$ parameters for the physical quantities. The difference between the general and the supersymmetric approaches becomes manifest only in the study of the LFV processes: we get
\beq
\frac{BR(\ell_i\to \ell_j\gamma)}{BR(\ell_i\to \ell_j\nu_i{\bar \nu_j})}= \frac{48\pi^3 \alpha}{G_F^2 M^4}\left[\vert w^{(1)}_{ij} u^2\vert^2+\frac{m_j^2}{m_i^2} \vert w^{(2)}_{ij} u\vert^2\right]\;,
\label{LFVsusy}
\eeq
where $w^{(k)}_{ij}$ are general coefficients of order 1.
Notice that now the first contribution on the right-hand side of eq.(\ref{LFVsusy}) is suppressed by a factor of $|u|^2$ compared to the non supersymmetric case of eq.(\ref{LFV}).
In most of the allowed range for $|u|$, the branching ratios of $\mu\to e \gamma$ and $\tau \to\mu \gamma$ are similar and larger than
the branching ratio of $\tau\to e \gamma$.
Assuming $\vert w^{(1,2)}_{\mu e}\vert=1$, the present (future) experimental limit on $BR(\mu\to e \gamma)$ implies the following bounds
\begin{eqnarray}
\vert u\vert =& 0.001&\qquad M>0.7~(2)~~{\rm TeV}\\
\vert u\vert =& 0.05&\qquad M>14~(48)~~{\rm TeV}\;.
\end{eqnarray}
We see that at variance with the non-supersymmetric case there is a range of permitted values of the parameter $|u|$
for which the scale $M$ can be sufficiently small to allow an explanation of the observed discrepancy in $a_\mu$,
without conflicting with the present bound on $BR(\mu\to e \gamma)$.
\begin{figure}[h]
\includegraphics[width=10cm]{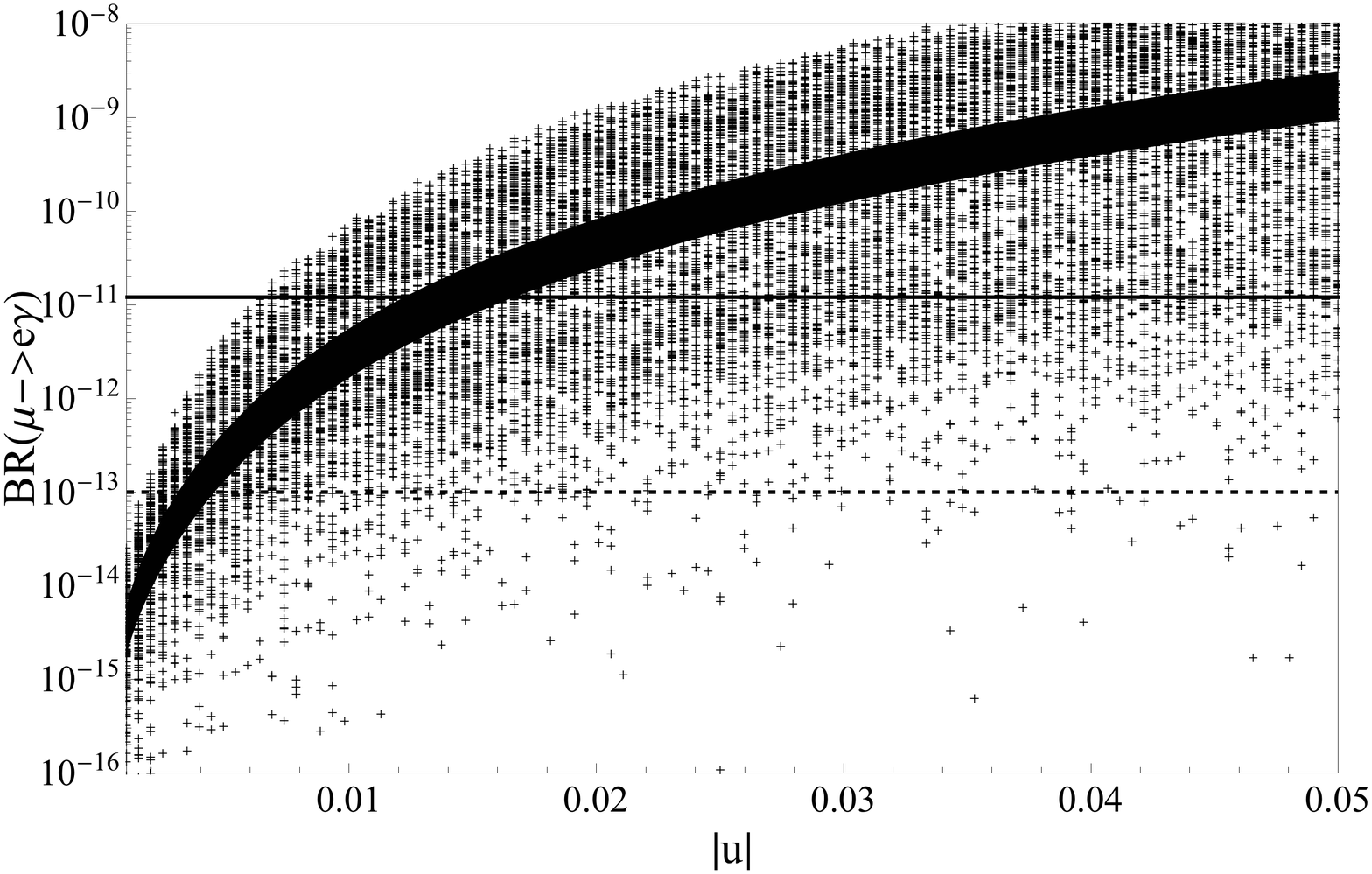}\hspace{2pc}
\begin{minipage}[b]{14pc}\caption{\label{BR}The branching ratio of $\mu\to e \gamma$ as a function of $|u|$, eq.(\ref{muegamma}). See the text for explanations.}
\end{minipage}
\end{figure}

We can eliminate the dependence on the unknown scale $M$ by combining eqs.(\ref{DM}) and (\ref{LFVsusy}). For $\mu\to e \gamma$
we get:
\begin{equation}
\frac{BR(\mu\to e\gamma)}{BR(\mu\to e\nu_\mu{\bar \nu_e})}= \frac{12\pi^3 \alpha}{G_F^2 m_\mu^4}\left(\delta a_\mu\right)^2
\left[\vert \tilde{w}^{(1)}_{\mu e}\vert^2 \vert u\vert^4+\frac{m_e^2}{m_\mu^2} \vert \tilde{w}^{(2)}_{\mu e}\vert^2\vert u\vert^2\right]
\label{muegamma}
\end{equation}
where $\tilde{w}^{(1,2)}_{\mu e}$ are unknown, order 1 coefficients. We plot $BR(\mu\to e\gamma)$ versus $|u|$ in fig. 1,
where the coefficients $\tilde{w}^{(1,2)}_{\mu e}$ are kept fixed to 1 (darker region) or are random
complex numbers with absolute values between zero and two (lighter region). The deviation of the anomalous magnetic moment
of the muon from the SM prediction is in the interval of the experimentally allowed values,
about three sigma away from zero. The continuous (dashed) horizontal line corresponds to the present (future expected) experimental bound on $BR(\mu\to e\gamma)$. Even if the ignorance about the coefficients $\tilde{w}^{(1,2)}_{\mu e}$ does not allow us
to derive a sharp limit on $|u|$, we see that the present limit on $BR(\mu\to e \gamma)$ disfavors values of $|u|$
larger than few percents. We recall that in this model the magnitudes of $|u|$ and $\theta_{13}$ are comparable and therefore we can derive an indication for the value of the reactor angle, which cannot be larger than few percents.


\section{Conclusions}
\label{section:Conclusions}

The lepton mixing matrix is well approximated by the TB mixing, which is easily recovered by flavour models based on the discrete group $A_4$: among the others, $A_4\times Z_3\times U(1)_{FN}$ represents a sort of minimal choice. We have studied the predictions of this model, both in a supersymmetric and in a general non supersymmetric context, for a set of observables and we can conclude that the supersymmetric version suggests the presence of NP at about a few TeV, which explains the discrepancy in $(g-2)_\mu$ and a probably positive signal for $\mu\to e\ga$ at MEG and indicates an upper bound for $\theta_{13}$ of few percents.


\ack
I thank the organizers of \emph{DISCRETE'08 - Symposium on Prospects in the Physics of Discrete Symmetries} for giving the opportunity to present my talk and for the kind hospitality in Valencia.
Alike I thank Ferruccio Feruglio, Claudia Hagedorn and Yin Lin for the pleasant and advantageous collaboration.


\end{document}